\documentclass{article}

\usepackage{graphicx,authblk}
\usepackage[margin=1.0in]{geometry}

\title{Membrane-less hydrogen bromine flow battery}
\author[1]{William A. Braff}
\author[2]{Martin Z. Bazant}
\author[1]{Cullen R. Buie\thanks{crb@mit.edu, +617 253 9379}}
\affil[1]{Department of Mechanical Engineering, Massachusetts Institute of Technology, Cambridge, MA}
\affil[2]{Department of Chemical Engineering, Massachusetts Institute of Technology, Cambridge, MA}

\begin{document}
	\maketitle

\begin{abstract}
In order for the widely discussed benefits of flow batteries for electrochemical energy storage to be applied at large scale, the cost of the electrochemical stack must come down substantially. One promising avenue for reducing stack cost is to increase the system power density while maintaining efficiency, enabling smaller stacks. Here we report on a membrane-less, hydrogen bromine laminar flow battery as a potential high power density solution. The membrane-less design enables power densities of 0.795 W cm$^{-2}$ at room temperature and atmospheric pressure, with a round-trip voltage efficiency of 92\% at 25\% of peak power. Theoretical solutions are also presented to guide the design of future laminar flow batteries. The high power density achieved by the hydrogen bromine laminar flow battery, along with the potential for rechargeable operation, will translate into smaller, inexpensive systems that could revolutionize the fields of large-scale energy storage and portable power systems.
\end{abstract}
\newpage

Low-cost energy storage remains a critical unmet need for a wide range of applications, include grid scale frequency regulation, load following, contingency reserves, and peak shaving, as well as portable power systems \cite{Yang:2011hy,Rugolo:2012td,Soloveichik:2011ea,Hittinger:2012hg}. For applications that require the storage of large quantities of energy economically and efficiently, flow batteries have received renewed attention \cite{Weber:2011eg,PoncedeLeon:2006dr}. A wide variety of solutions have been proposed, including zinc-bromine and vanadium redox cells \cite{Yang:2011hy,Knehr:2012br}. This includes recent efforts to incorporate novel concepts such as organic electrolytes for greater voltage stability and semisolid reactants for higher reactant energy density \cite{Liu:2010co,Duduta:2011cp} or chemistries to reduce reactant cost~\cite{SkyllasKazacos:1991vl,Wang:2012ff}. One such permutation is the hydrogen bromine flow battery \cite{Kosek:1988wk,Yeo:1980ub,Livshits:2006fk,Kreutzer:2012hb,Cho:2012bl}. The rapid and reversible reaction kinetics of both the bromine reduction reaction and the hydrogen oxidation reaction minimize activation losses \cite{Cooper:1970tb,GoorDar:2012et}, while the low cost (\$1.39 kg$^{-1}$) and abundance (243,000 metric tons produced per year in the United States alone) of bromine distinguishes it from many other battery chemistries \cite{NationalMineralsInformationCenter:2011tv}. However, theoretical investigations of such systems have revealed that the perfluorosulfonic acid membranes typically used suffer from low conductivity in the absence of sufficient hydration \cite{Clark:2012fy,Kreuer:2004fm}. In the presence of hydrobromic acid, this membrane behavior is the dominant limitation on overall performance \cite{Savinell:1988tn,Kreutzer:2012hb}.

Laminar flow electrochemical cells have been proposed to address many of the challenges that face traditional membrane-based systems \cite{Ferrigno:2002cl,Choban:2004jp,Jayashree:2005bj,Cohen:2005ed,Kjeang:2007cp,Kjeang:2008bc,Brushett:2009jl}. Laminar flow cells eliminate the need for an ion exchange membrane by relying on diffusion to separate reactants. Eliminating the membrane decreases cost, relaxes hydration requirements, and opens up the possibility for a much wider range of chemistries to be investigated. This flexibility has been exploited in the literature; examples include vanadium redox flow batteries \cite{Ferrigno:2002cl,Kjeang:2007cp}, as well as methanol \cite{Jayashree:2005bj}, formic acid \cite{Choban:2004jp}, and hydrogen fuel cells \cite{Brushett:2009jl}. However, none of these systems have achieved power densities as high as their membrane-based counterparts. This is largely because the proposed chemistries already work well with existing membrane technologies that have been refined and optimized over several decades. More recently, a laminar flow fuel cell based on borohydride and cerium ammonium nitrate employed a porous separator, chaotic mixing, and consumption of acid and base to achieve a power density of 0.25 W cm$^{-2}$ \cite{Mota:2012dr}. This appears to be the highest previously published power density for a membrane-less laminar flow fuel cell.

In this work, we present a membrane-less Hydrogen Bromine Laminar Flow Battery (HBLFB) with reversible reactions and a peak power density of 0.795 W cm$^{-2}$ at room temperature and atmospheric pressure. The cell uses a membrane-less design similar to previous work \cite{Jayashree:2005bj}, but with several critical differences that allow it to triple the highest previously reported power density for a membrane-less electrochemical cell and also enable recharging. First, where many previous laminar flow electrochemical cell designs were limited to low current operation by the low oxygen concentration at the cathode, the HBLFB uses gaseous hydrogen fuel and aqueous bromine oxidant \cite{Ferrigno:2002cl,Choban:2004jp,Jayashree:2005bj,Cohen:2005ed,Brushett:2009jl}. This allows for high concentrations of both reactants at their respective electrodes, greatly expanding the mass transfer capacity of the system. Next, both reactions have fast, reversible kinetics, with no phase change at the liquid electrode, eliminating bubble formation as a design limitation. These two characteristics of the HBLFB enable high power density storage and discharge of energy at high efficiency, while avoiding the cost and reliability issues associated with membrane-based systems.

\section{Results}
\subsection{Numerical model}
\begin{figure}
\includegraphics[width=9 cm]{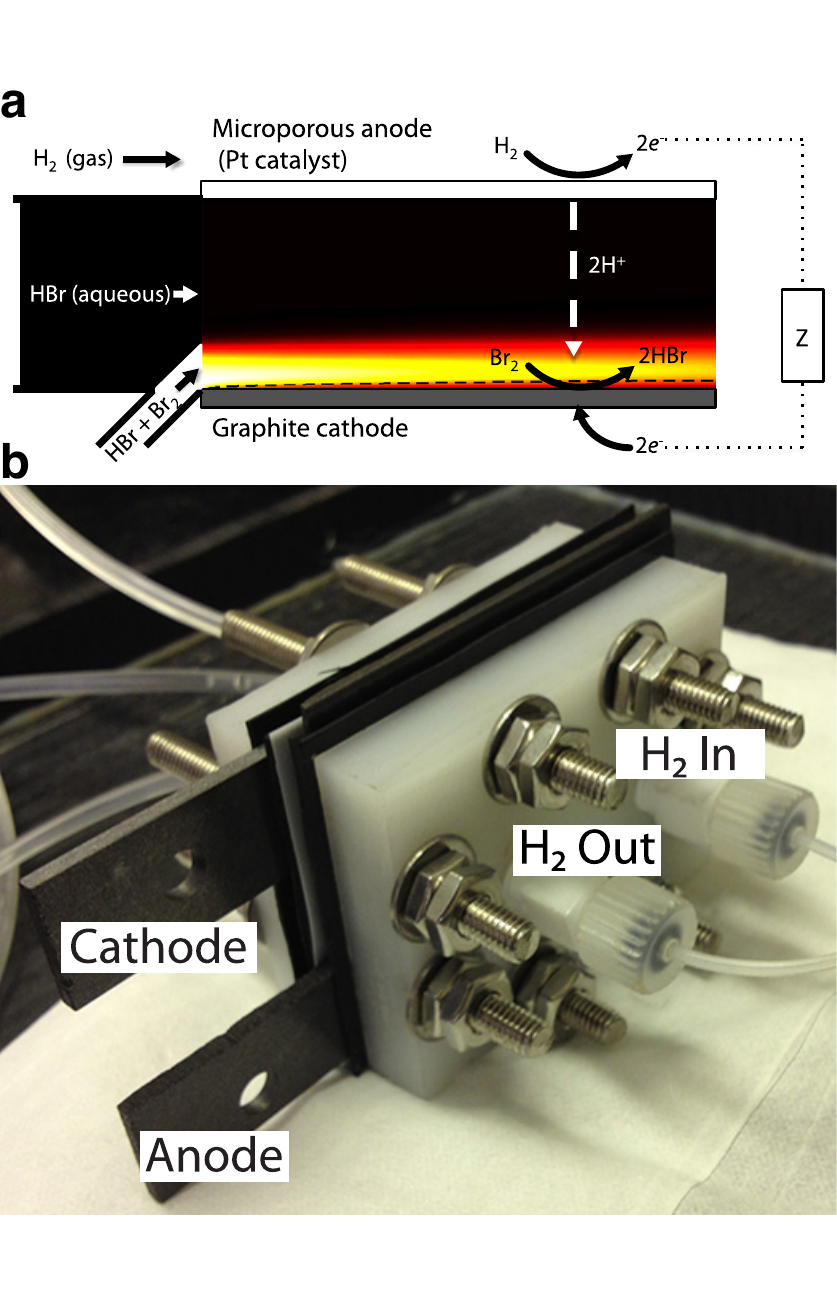}
  \caption{Schematic of reactant flow within the Hydrogen Bromine Laminar Flow Battery (HBLFB). During discharge, liquid bromine is reduced to hydrobromic acid along the lower solid graphite electrode, and hydrogen is oxidized at the upper porous electrode. Numerically predicted concentration of Br$_2$ for a Peclet number of 10,000 and initial concentrations of bromine and hydrobromic acid of 1 M is overlaid (a). Assembled cell prior to testing (b).}
  \label{fig:cell_design}
\end{figure}

A two-dimensional numerical model of the device was constructed to allow for a greater understanding of the underlying physics of the cell \cite{Braff:ib,Braff:2013kk}. The model solves the Nernst-Planck equations with advection in the imposed flow for the concentrations and electrostatic potential, assuming a quasi-neutral bulk electrolyte. Fully developed Poiseuille flow was assumed in the channel, with reactions occurring at the top and bottom of the channel along thin electrodes (Fig.~\ref{fig:cell_design}). Because the channel width far exceeds the height, edge effects can be ignored, validating the two-dimensional assumption. Equilibrium potentials along the cathode and anode were determined by the Nernst equation assuming dilute solution theory, and activation losses were estimated using the symmetric Butler-Volmer equation, consistent with existing kinetics data \cite{Cooper:1970tb}. Electrolytic conductivity was assumed to depend on the local (spatially evolving) hydrobromic acid concentration, and was calculated using empirical data \cite{Lide:2012ut}. 

\subsection{Discharge experiments}
\begin{figure}
\includegraphics[width=9 cm]{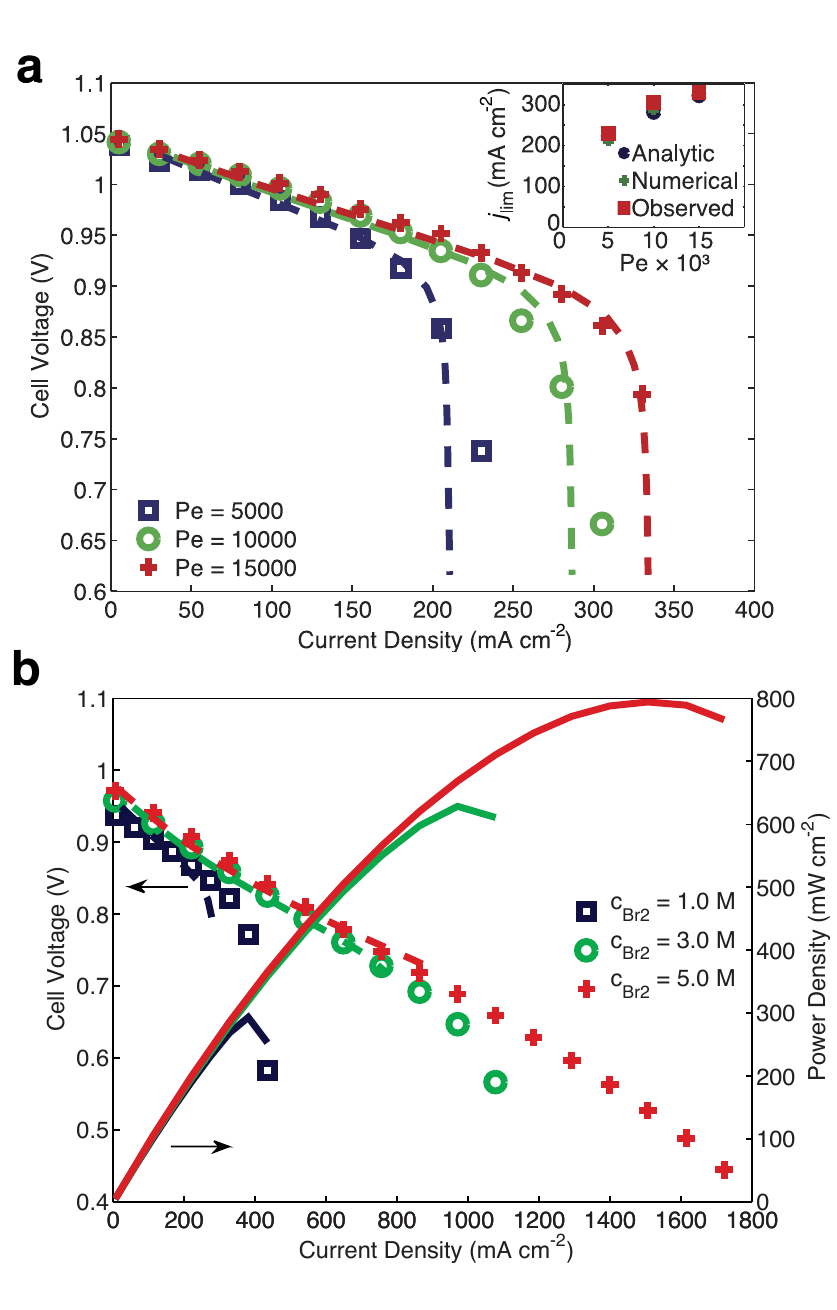}
  \caption{Discharge performance of the HBLFB. Predicted (dashed) and observed (symbols) cell voltage as a function of current density and Peclet number for the HBLFB during discharge using 1M Br$_2$ and 1M HBr. For low concentration reactants, mass transport is the dominant source of loss, and limiting current density $j_{lim}$ can be predicted analytically as a function of Peclet number $Pe$ using the L\'ev\^eque approximation (inset) (a). Predicted (dashed) and observed (symbols) cell voltage and power density as a function of current density and $Br_2$ concentration for $Pe = 10,000$ and 3 M HBr. The higher reactant concentrations require concentrated solution theory to accurately model, but allow for much higher current and power density (b).}
  \label{fig:pol_curve}
\end{figure}

The cell was operated galvanostatically at room temperature and atmospheric pressure over a range of flow rates and reactant concentrations. The cell was observed to reach steady state in less than ten seconds, so each data point was collected after sixty seconds of steady state operation to eliminate transient artifacts. Polarization data was collected as a function of Peclet number for the HBLFB using 1 M HBr and 1 M Br$_2$, and compared with numerical model results (Fig.~\ref{fig:pol_curve}). The Peclet numbers of 5,000, 10,000, and 15,000 correspond to Reynolds numbers of 5.75, 11.5, and 17.25, oxidant flow rates of 0.22, 0.44, and 0.66 mL min$^{-1}$ cm$^{-2}$, and mean velocities of 6.3, 12, and 19 mm s$^{-1}$, respectively. The oxidant flow rates correspond to stoichiometric currents of 0.7, 1.4, and 2.1 A cm$^{-2}$ respectively. Hydrogen was flowed in excess at a rate of 25 sccm. The slightly enhanced maximum current density of the observed results compared to the predicted results may be attributed to the roughness of the electrode surface producing chaotic mixing that slightly enhances reactant transport. Below limiting current, the agreement between model and experiment is very good. At low current densities, the voltage differences between the different Peclet numbers are small, but in each case, the voltage drops rapidly as the cell approaches limiting current, corresponding to the predicted mass transfer limitations. 

\subsection{Analytical limiting current} 
These limitations can be estimated analytically for fast reactions by assuming a bromine concentration of zero at the cathode and applying the L\'ev\^eque approximation to the bromine depletion layer \cite{Leveque:1928vb,Probstein:2005uq}. For a channel of length $L$ and height $h$, fully developed Poiseiulle flow, and initial bromine concentration $c_0$, the resulting partial differential equation and boundary conditions can be expressed in terms of dimensionless bromine concentration $\tilde{c} = c/c_0$, position $\tilde{y} = y/h$ and $\tilde{x} = x/h$, channel aspect ratio $\beta = L/h$, and Peclet number $\textrm{Pe} = Uh/D$.
\begin{eqnarray}
\frac{6\textrm{Pe}}{\beta}\tilde{y}\frac{\partial\tilde{c}}{\partial \tilde x} &=& \frac{\partial^2 \tilde{c}}{\partial \tilde{y}^2}\\
\tilde{c}(\tilde{x}=0) &=& 1\nonumber\\
\tilde{c}(\tilde{y}\rightarrow\infty) &=& 1\nonumber\\
\tilde{c}(\tilde{y}=0) &=& 0\nonumber
\end{eqnarray}
A similarity technique can be applied to convert this to an ordinary differential equation. 
\begin{eqnarray}
\eta^2\frac{\partial \tilde{c}}{\partial \eta} &=& \frac{\partial^2\tilde{c}}{\partial \eta^2}\\
\tilde{c}(\eta=0) &=& 0\nonumber\\
\tilde{c}(\eta\rightarrow\infty) &=& 1\nonumber\\
\eta &=& \frac{\tilde{y}}{\sqrt[3]{\beta\tilde{x}/2\textrm{Pe}}}\nonumber
\end{eqnarray}
This equation can be solved exactly in terms of the incomplete Gamma function, $\Gamma(s,a)$.
\begin{eqnarray}
\tilde{c}(\tilde{x},\tilde{y}) &=& \Gamma\left(\frac{2\textrm{Pe}\tilde{y}^3}{3\beta\tilde{x}},\frac{1}{3}\right)
\end{eqnarray}

Limiting current can be calculated using Faraday's law to determine the distribution of current along the length of the electrode and integrating to obtain limiting current $j_{\textrm{lim}}$ as a function of Peclet number $\textrm{Pe}$, reactant concentration $c_0$ and diffusivity $D$, channel height $h$ and aspect ratio $\beta$, Faraday's constant $F$, and the number of moles of electrons transferred per mole of reactant $n$.
\begin{eqnarray}
j_{\textrm{lim}} &=& 1.47\frac{c_0nDF}{h} \sqrt[3]{\frac{\textrm{Pe}}{\beta}}
\end{eqnarray}
This result has considerable bearing on how laminar flow systems should be designed and operated. The presence of aspect ratio in the denominator means that shortening the channel results in greater power density, as observed in experiment \cite{Thorson:2012fg}. In addition, the weak 1/3 dependence on Peclet number means that increasing the flow rate beyond a certain point yields minimal benefits. There was excellent agreement between maximum observed current density, maximum numerically predicted current density, and the analytically predicted limiting current density as a function of Peclet number (fig.~\ref{fig:pol_curve}a).

\subsection{High power operation}
Higher bromine concentrations were also investigated by using a more concentrated electrolyte (3M HBr) to enhance bromine solubility and move beyond the mass transfer limitations of 1 M Br$_2$. The performance of this system was investigated at a Peclet number of 10,000 by varying the bromine concentration (fig.~\ref{fig:pol_curve}b). Using 5M Br$_2$ and 3M HBr as the oxidant and electrolyte respectively, a peak power density of 0.795 W cm$^{-2}$ was observed when operated near limiting current density. This corresponds to power density per catalyst loading of 1.59 W mg$^{-1}$ platinum. The open circuit potential of the cell dropped more than might be predicted simply using the Nernst equation at the cathode. The drop in open circuit potential is consistent with data on the activity coefficient of concentrated hydrobromic acid available in the literature, as well as previous studies that employed concentrated hydrobromic acid \cite{Lide:2012ut,Livshits:2006fk}. To account for this effect, empirical data for the activity coefficient of hydrobromic acid as a function of local concentration was incorporated into the boundary conditions. The activity coefficient was assumed to vary slowly enough within the electrolyte that gradients in activity coefficient were neglected in the governing equations. Taking into account the activity coefficient of hydrobromic acid, there is good agreement between the model and experiment. As in the low concentration data, the observed maximum current density is slightly higher than predicted. Otherwise the model captures the main features of the data, including the transition from transport limited behavior at low bromine concentrations, evidenced by a sharp drop in cell voltage, to ohmically limited behavior at high bromine concentrations. In this ohmically limited regime, mass transport limitations are less important, and the limiting current solution applied to the low concentration results does not apply.

\subsection{Recharging and round-trip efficiency}
\begin{figure}
\includegraphics[width=9 cm]{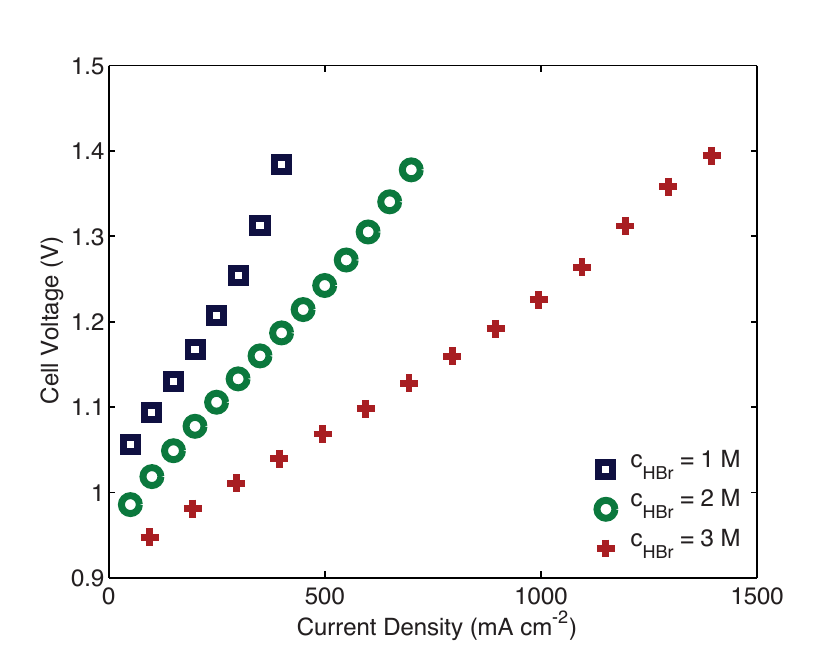}
  \caption{Charging performance of the HBLFB. Observed cell voltage during charging as a function of HBr concentration at a Peclet number of 10,000 and a Br$_2$ concentration of 1 M. Increasing the HBr concentration increases both the conductivity and the limiting current, resulting in superior performance.}
  \label{fig:charging_data}
\end{figure}

\begin{figure}
\includegraphics[width=9 cm]{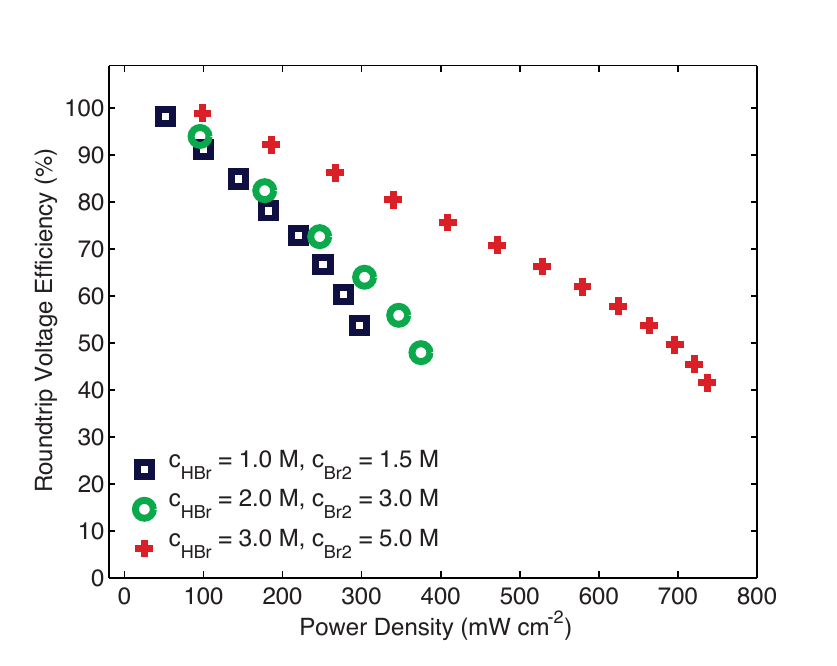}
  \caption{Roundtrip efficiency of the HBLFB. Round-trip voltage efficiency of the HBLFB as a function of power density for a range of reactant concentrations at a Peclet number of 10,000.}
  \label{fig:power_efficiency}
\end{figure}

Charging behavior was also investigated by flowing only HBr and applying a voltage to electrolyze HBr back into Br$_2$ and H$_2$. The voltage versus current density behavior of the cell was investigated during charging as a function of HBr concentration at a Peclet number of 10,000 (fig.~\ref{fig:charging_data}). Experimental conditions were kept identical to those of the discharge experiments, with the exception that no bromine was externally delivered to the cell. Side reactions, in particular the formation of hypobromous acid and the evolution of oxygen become dominant before potentials get sufficiently high to observe limiting current, the numerical model cannot accurately describe this behavior at potentials above 1.3 volts, and was not applied to the charging data. At lower voltages, both the electrolyte conductivity and the limiting current density were increased by increasing the HBr concentration, resulting in increased performance (fig.~\ref{fig:charging_data}). Roundtrip voltage efficiency was then calculated by dividing the discharge voltage by the charging voltage for a number of reactant concentrations (fig.~\ref{fig:power_efficiency}). Voltage efficiencies slightly greater than 100\% were observed for low power densities due to differences in the open circuit potential that are generated by the variation in bromine concentration between the charging and discharging experiments, but this anomaly becomes unimportant at higher current densities, where the reactant concentration varies spatially much more strongly. Using high concentration reactants, roundtrip efficiency of 90\% was observed when using high concentration reactants at 25\% of peak power (200 mW/cm$^2$ during discharge). This appears to be the first publication of roundtrip charging and discharging of a membrane-less laminar flow battery, and compares very favorably to existing flow battery technologies. Vanadium redox batteries, for example, have demonstrated voltage efficiencies as high as 91\%, but only at a discharge power density nearly an order of magnitude lower than the HBLFB~\cite{HaddadiAsl:1995wh}. 

Although closed-loop operation was not demonstrated in this study, some insight into coulombic efficiency can be gathered by considering the effect of reactant mixing within the channel. At 25\% of peak power, the single-pass coulombic efficiency of the cell is only about 1\%. If no attempt were made to separate the reactants at the outlet of the cell, the resulting energy efficiency of the cell would also be very low. However, if the channel outlet were split to conserve the volume of stored electrolyte and oxidant, only the Br$_2$ that had diffused into the electrolyte layer would be lost. Assuming fully developed Poiseuille flow, this corresponds to a coulombic efficiency of 72\%, or a roundtrip energy efficiency of 66\%.

\section{Discussion}
Several opportunities for improvement on this design remain to be investigated. First, the porous hydrogen anode was selected because of its commercial availability, despite the fact that it was intended for use in a proton exchange membrane fuel cell. The catalyst composition and structure, wettability, and media porosity have not been optimized. Previous work has shown that these parameters can impact the power density of laminar flow cells by nearly an order of magnitude \cite{Brushett:2012bj}. Recent work on thin hydrogen oxidation electrodes has demonstrated that excellent catalytic activity can be achieved with platinum loadings as low as 3 $\mu$g cm$^{-2}$, more than two orders of magnitude lower than the electrodes used in this study \cite{Wesselmark:2010ij}. Assuming equivalent performance, an HBLFB employing such a hydrogen electrode would have a specific power density of roughly 250 W mg$^{-1}$ platinum, virtually eliminating platinum as a cost-limiting component of the system. Second, the channel geometry used in these experiments was relatively long in order to achieve high oxidant utilization. Shortening the channel would decrease the average thickness of the depletion layer that develops along the cathode, enabling higher current densities \cite{Thorson:2012fg}. Third, reducing the distance between electrodes would greatly reduce ohmic losses, which are dominant when high concentration reactants are fed into the cell. Incorporating further refinements, such as micro patterned chaotic mixing patterns coupled with a non-specific convection barrier or non noble metal catalysts could also improve performance. \cite{Stroock:2002hq,Mota:2012dr,Liu:2012aa}.

The initial data presented here for the HBLFB suggests that high power density and high efficiency energy storage is achievable using a membrane-less electrochemical cell operating at room temperature and pressure. The HBLFB requires no special procedures or facilities to fabricate, and uses kinetically favorable reactions between abundant, low cost reactants. Recent work has shown that a membrane-based hydrogen-bromine flow battery at room temperature can generate 850 mW cm$^{-2}$, or 7\% more power than these experiments with the HBLFB at room temperature \cite{Cho:2012bl}. However, this was achieved using a stoichiometric oxidant flow rate over 8 times larger than that used in this work, as well as an acid-treated porous bromine electrode with substantially greater active area than the bare graphite bromine electrode of the HBLFB. 

This work represents a major advance of the state of the art in flow batteries. To the best of the authors' knowledge, the data presented here represent the highest power density ever observed in a laminar flow electrochemical cell by a factor of three, as well as some of the first recharging data for a membrane-less laminar flow electrochemical cell \cite{Mota:2012dr,Lee:2013ji}. Although previous work has identified the appropriate scaling laws \cite{Ismagilov:2000ky,Bazylak:2005br}, the result presented here represents the first exact analytical solution for limiting current density applied to a laminar flow electrochemical cell, and serves as a guide for future designs. The HBLFB rivals the performance of the best membrane-based systems available today without the need for costly ion exchange membranes, high pressure reactants, or high temperature operation. This system has the potential to play a key role in addressing the rapidly growing need for low-cost, large-scale energy storage and high efficiency portable power systems.

\section{Methods}
\subsection{Cell fabrication}
A proof of concept electrochemical cell was assembled using a graphite cathode and a commercial carbon cloth gas diffusion anode with 0.5 mg cm$^{-2}$ of platinum (60\% supported on carbon) obtained from the Fuel Cell Store (San Diego, CA). The cell was housed between graphite current collectors and polyvinylidene fluoride (PVDF) porting plates (fig.~\ref{fig:cell_design}b). All components were fabricated using traditional CNC machining or die cutting.  No additional catalyst was applied to the cathode. The hydrogen flow rate through the porous anode was 25 sccm, and an oxidant stream of bromine mixed with aqueous hydrobromic acid passed through the channel in parallel with an electrolyte stream of aqueous hydrobromic acid. An 800 $\mu$m thick Viton gasket was used to separate the two electrodes and create the channel, which was 1.4 cm long from oxidant inlet to outlet with an active area of 25 mm$^2$. A fixed ratio of ten to one was maintained between the electrolyte and oxidant flow rates, and the net flow rate was adjusted to study the performance of the cell as a function of the Peclet number, $\textrm{Pe}$, defined by the average flow velocity $U$, channel height $h$, and bromine diffusion coefficient $D$ such that $\textrm{Pe} \equiv Uh/D$.

\subsection{Numerical model details}
A scaled, dimensionless model was constructed in COMSOL Multiphysics, and results were calculated over a range of flow rates and reactant concentrations. A complete description of the model has been presented previously \cite{Braff:ib,Braff:2013kk}. Bromine concentration varied along the length of the channel, resulting in strong spatial variations in the current density (fig.~\ref{fig:cell_design}a). Current-voltage data was obtained for a range of reactant concentrations and flow rates for comparison to experimental data by averaging the current density along the length of the cell, and calculating solutions over a range of specified cell voltages.

\section{Acknowledgements}

The authors acknowledge financial support from the Department of Defense (DoD) through the National Defense Science \& Engineering Graduate Fellowship (NDSEG) Program, as well as the MIT Energy Initiative Seed Fund.

\section{Author Contributions}
W.A.B. performed the experiments and simulations; M.Z.B. developed the theoretical plan, and C.R.B. the experimental plan. All authors wrote and edited the manuscript.

\section{Competing Financial Interests Statement}
The authors have no competing financial interests.

\end{document}